# Collisions between cold molecules in a superconducting magnetic trap

Yair Segev*, Martin Pitzer*, Michael Karpov*, Nitzan Akerman, Julia Narevicius and Edvardas Narevicius

Department of Chemical and Biological Physics, Weizmann Institute of Science, Rehovot 7610001, Israel.

**Abstract**

Collisions between cold molecules are essential for studying fundamental aspects of quantum chemistry, and may enable formation of quantum degenerate molecular matter by evaporative cooling. However, collisions between trapped, naturally occurring molecules have so far eluded direct observation due to the low collision rates of dilute samples. We report the first directly observed collisions between cold, trapped molecules, achieved without the need of laser cooling. We magnetically capture molecular oxygen in a 0.8K·$k_B$ deep superconducting trap, and set bounds on the ratio between the elastic and inelastic scattering rates, the key parameter determining the feasibility of evaporative cooling. We further co-trap and identify collisions between atoms and molecules, paving the way to studies of cold interspecies collisions in a magnetic trap.

**Introduction**

Cold bimolecular collisions play a central role in a variety of processes, including interstellar chemistry[1], and are used for the detailed examination of fundamental physics in two-body interactions[2–5] and the preparation of new states of quantum matter[6]. Experimental efforts started with the pioneering work of Herschbach and Lee[7], with the first use of a crossed molecular beam apparatus for physical chemistry. State of the art experiments in moving frames of reference are ever pushing the limits to lower collision energies[8–12], but the low collision rates associated with such cold, dilute gases eventually necessitate trapping for long durations.

Laser cooling, the workhorse of the cold atoms field, has enabled the production of designer molecules in traps via magneto-association of cold atoms. The creation of such Feshbach molecules provided the groundbreaking observation of the BCS-BEC crossover[13], as well as ultracold reactions[2,3] and collisions[14,15] between the molecules. In the last few years, laser cooling has been applied directly to molecules in magnetic and magneto-optical traps using sophisticated schemes, bringing molecules to the microkelvin regime for the first time[16–19]. A recent work reports the first observation of collisions between laser-cooled molecules in optical tweezers[20].

Most molecules found in nature, however, have no suitable electronic transitions for the application of laser cooling. There have thus been many efforts to observe collisions between trapped molecules without the use of laser-cooling, starting with buffer gas cooled magnetically trapped CaH[21] and electrostatically trapped ammonia[22], OH[23,24], OD[25] and NH radicals[26], followed by magnetically trapped $CH_3$[27]. In other approaches, dilute molecular samples have been co-trapped with laser or buffer gas cooled atoms[28–30]. Electrostatic trapping of OH radicals[31] enabled observation of Majorana transition enhancement[32]. Despite these efforts, trapped samples with sufficient density to allow direct observation of bimolecular collisions of such species have not been achieved.

* Corresponding authors: yair.segev@weizmann.ac.il (Y.S.); edn@weizmann.ac.il (E.N.)          Feb. 2019

Here we present the first directly measured collisions between cold, trapped molecules achieved without any laser cooling. We obtain a high-density ensemble of molecules by adiabatically decelerating oxygen using co-moving magnetic traps. The molecules are loaded into a stationary magnetic trap, formed by high-temperature superconducting (HTS) coils carrying high DC currents, and held for up to 90 seconds. The combination of high density and long interaction time enables observing two-body collisions between molecules by measuring a clear non-exponential decay of the trapped molecule number. We provide an independent proof of sufficiently high collision rates in the trap by co-loading a dilute sample of lithium atoms alongside the higher density molecules. Such a capability will enable future studies of ultracold reactions between co-trapped paramagnetic species.

A high collision rate is a prerequisite for evaporative cooling towards the ultracold regime. We use the robustness of the HTS trap to compress the molecular sample using transient fields on the order of hundreds of Tesla per second, a step routinely applied to increase the collision rates in atomic samples[33]. This capability will also enable collisional studies with other molecular radicals that are generated in lower numbers compared to molecular oxygen. Finally, we make use of the trap's tunable depth and the spatial resolution of our detection method to characterize the collisional properties of cold oxygen molecules and place bounds on the ratio between elastic to inelastic scattering rates, a key parameter controlling the evaporative cooling mechanism.

**Experimental method**

Our choice of particles are paramagnetic molecules and atoms, which can be trapped with magnetic fields by making use of the Zeeman effect. Practical field strengths, on the scale of 1T, are sufficient to trap particles at typical velocities of a few to several tens of meters per second, equivalent to a thermal distribution at around 1K. To create an ensemble with sufficiently low mean velocity as well as a narrow distribution, we load our trap with an adiabatically decelerated bunch of particles cooled by supersonic expansion.

The molecules are first emitted from a pulsed Even-Lavie valve[34], from which they expand and cool, in the translational as well as internal degrees of freedom[35]. The velocity of the ensuing beam is lowered by premixing the molecular gas into a heavier carrier gas, as well as by cooling the valve. The axial portion of the expanded plume enters the decelerator tube. Here, a sequence of timed pulses of current through pairs of coils creates a co-moving magnetic quadrupole field around a slice of the beam, which acts as a three-dimensional trap for the low-field seeking fraction of the particles. An asymmetry in the restoring force generated by the coil switching sequence creates an average axially decelerating force. A high fraction of particles remain in the moving trap if the deceleration profile is adiabatic[36,37].

The particles emerge from the decelerator at a velocity of several meters per second and enter the adjacent stationary trap. For this trap, we use coils made of superconducting wires[38], which can sustain high DC current densities that enable large magnetic field gradients on the order of 1T/mm. Our coils have a bore diameter of 7mm, an outer diameter of 17mm, and a width of 2mm. A disc of permendur is placed between the coils to increase the radial magnetic field gradient. The maximum constant current densities exceed 500A/mm$^2$, and produce DC fields of about 1T at the trap edge.

Loading the trap requires strong transient fields, as the particles would transit the trap in less than a millisecond. In these conditions, the critical current of a superconductor can be significantly lower than its DC value. Low temperature superconductors have previously been used for magnetically trapping atoms [39] and molecules [21]. We found that tape made of high temperature superconductors sustained higher transient fields than wires of low-temperature superconductors with comparable critical DC current densities. Cooling the HTS coils to 15K, far below their transition temperature $T_c$ of ~90K, allow us to turn on the trap to its full depth within several hundred microseconds without quenching.

After the loading sequence is completed, the currents to the trap are switched from a fast-rise capacitor bank to a high current DC power supply. From this point, the current can be maintained at a constant value, or varied in time to induce compression or expansion of the trapped molecular cloud. In the present design, heat dissipation from the current leads inside the vacuum limits the trapping duration at the highest currents. The density of particles in the trap at a given time is probed by laser ionization, where an electrostatic field is used to extract the ions from the trap onto a micro-channel plate (MCP).

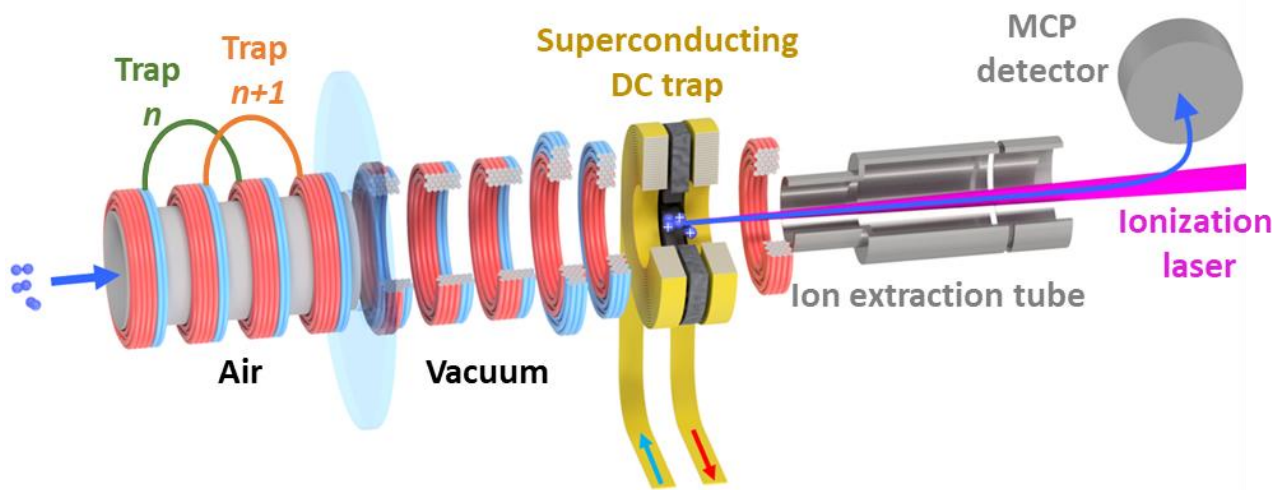

**Figure 1**: Experimental setup. Cold molecules emitted from a supersonic source enter the decelerator tube (left), and are adiabatically slowed by 480 co-moving magnetic traps, comprised of asymmetric pairs (red and blue) of pulsed coils. The molecules are loaded in between two high-temperature superconducting coils, where high DC currents are ramped within 0.5ms to form a stationary trap. After the desired trapping time, the density in the trap is probed using an ionization laser, with ions extracted to an MCP detector. Moving the laser focal point in the transverse plain provides spatial resolution of the trapped density.

## Results

We chose to first test the trapping scheme with molecular oxygen. Our decelerator can bring oxygen to a stop from 375m/s when it is emitted as a mixture with krypton carrier gas, and with the valve cooled to 165K [37]. Our deepest static trap for oxygen was about 800mK, as calculated by finite element analysis for an 80A DC current. Shallower traps are achieved by reducing the current after loading.

Detection is carried out using a 2+1 REMPI scheme[40], with a 225nm pulse of about 3mJ pulse energy and 5 ns duration focused by an f=400 mm lens into the trap. The measurements are sensitive to radial movement of the focal point within the trap, but insensitive to axial movement, indicating that the measurement is a column-integration of density along the laser beam axis, with a radius substantially smaller than the trap radius of 3mm.

Loading into a shallow trap of 50mK, we observe trapped $O_2$ for more than 90 seconds (Fig. 2 inset). The signal decays exponentially with a lifetime of $\tau \sim 52$s, consistent with our estimated background collision rate and indicating the absence of intermolecular collisions. For higher trap depths, however, we observe both an increase in initial signal as well as a clear deviation from an exponential decay, which becomes more pronounced at higher trap depths (Fig. 2).

For short trapping times up to five seconds, when losses due to background collisions are negligible, we can model the measured decay of the column-integrated density signal $Y$ by the two-body loss rate equation,

$$dY/dt = -aY^2.$$

For the deepest trap of 800mK the estimated two-body lifetime, defined as in Ref[15] as $\tau_{2B}=(aY_{(t=0)})^{-1}$, is approximately 9 seconds. This value is significantly shorter than the background limited lifetime, but orders of magnitude longer than the characteristic time scale of the trap dynamics. The two-body collision mechanisms that comprise this depletion may include inelastic contributions, such as spin changing collisions due to spin-rotation interaction[41], as well as elastic losses due to high velocity collisions that lead to evaporation of particles from the trap.

Two-body collisions between trapped species can be independently identified as the main loss mechanism by adding a more dilute species to the trap. We load and trap lithium atoms by laser ablating a solid target near the valve, and decelerating the atoms with the same co-moving traps[37]. The same wavelength used for detecting $O_2$ is also efficient for single-photon ionization of lithium.

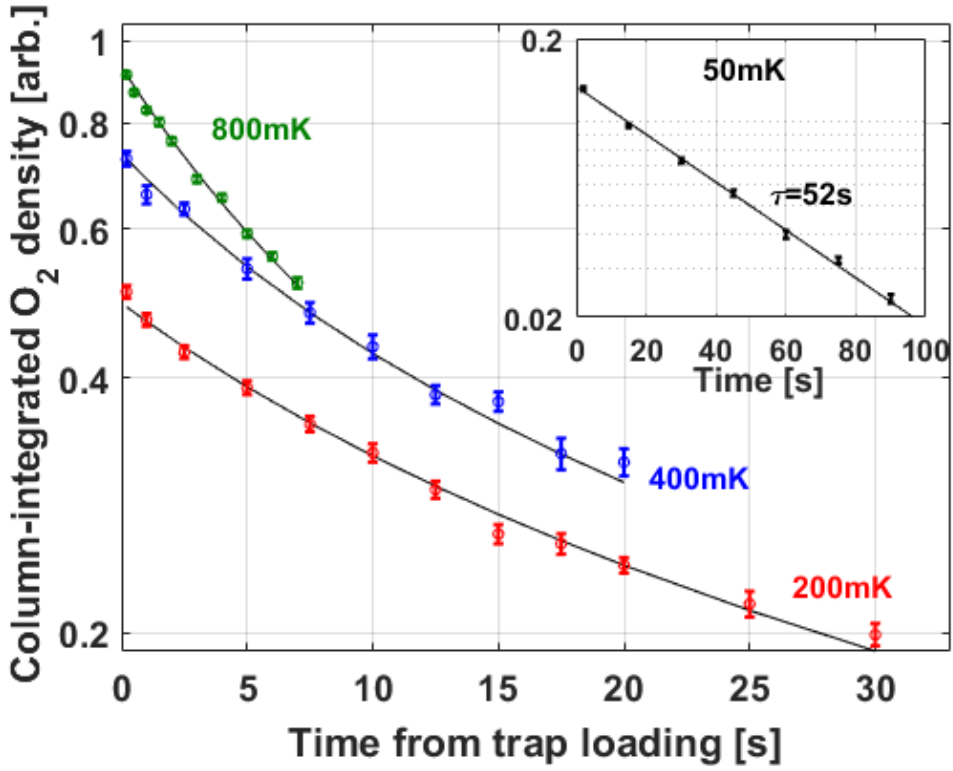

**Figure 2:** Column integrated density signal through the trap center for different trap depths. Trapped oxygen is observed in a shallow trap for 90s, with a 52s background-collision limited lifetime (inset). Deeper traps exhibit a higher initial signal, which decays non-exponentially, indicating two-body losses in the trap. Solid lines represent fits for a loss model based on bimolecular collisions.

When lithium is loaded into a trap without molecules, using a carrier beam of pure krypton, we observe an exponential decay with a lifetime of about $\tau$=14 seconds (Fig. 3). This rate, which is faster than the $O_2$ background-collision limited lifetime, is consistent with the larger cross-section for lithium collisions with the background $H_2$ due to larger electric-dipole polarizability of lithium compared to $O_2$ [37]. When we co-load lithium with $O_2$, we observe a fast, non-exponential decay of the atoms from the trap, indicating collisions with the trapped molecules. We find that this decay fits well with a two-body loss rate equation based on collisions between oxygen and lithium,

$$dY_{Li}/dt = -aY_{Li}Y_{O2}.$$

Here, the two-body lifetime of lithium increases as the oxygen is depleted. At early times we find a lifetime of $\tau_{2B,Li}=(aY_{O2(t=0)})^{-1}$=1.7s, about five times shorter than the oxygen two-body lifetime. A possible reason for this faster depletion is that lithium is more susceptible to evaporative losses following elastic collisions with $O_2$, due to its lighter mass which leads to higher reflected velocities, as well as the lower trap depth it experiences as a consequence of its lower magnetic moment. Note that the oxygen lifetime did not show any change due to the addition of lithium, validating our assessment that the lithium density is much lower than that of oxygen[37].

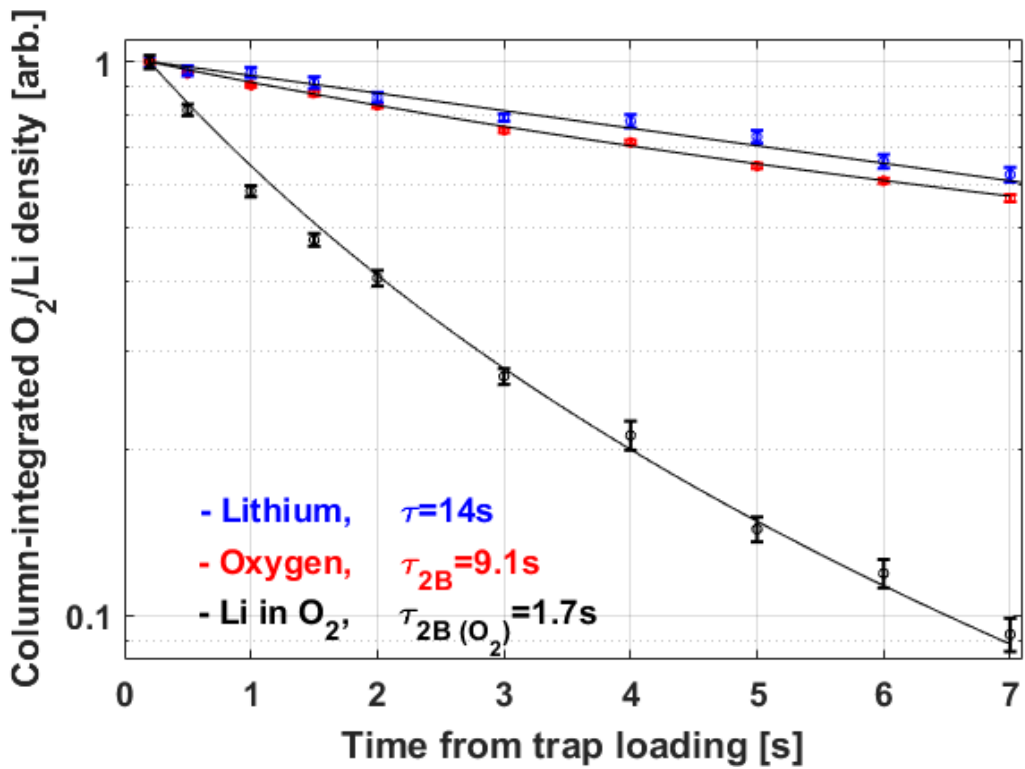

**Figure 3:** Trapping of lithium. When lithium is trapped without molecules we observe a 14s lifetime exponential decay, consistent with the estimation of the background-collision limit. When lithium is trapped alongside oxygen, the Li signal exhibits a fast decay indicating atom-molecule collisions. The solid lines represent fits for the respective loss models: exponential decay of lithium alone due to background collisions; two-body decay of oxygen due to bimolecular collisions; and two-body decay of lithium due to collisions with oxygen molecules, with time-dependent oxygen density.

Probing different portions of the trap can provide further information regarding the mechanisms of trap depletion. To locate the trap center, we load oxygen into a 50mK trap and adiabatically ramp the trap depth within 20ms to 800mK, thus compressing the trap volume, and sample the signal before and after the ramp. When the laser passes through the trap center, we observe a sharp increase in signal after the ramp. In contrast, probing near the circumference of the compressed trap we see an almost complete loss of signal, as the particles are pushed inwards during compression (Fig. 4A).

Once the trap center is clearly discerned, we measure the evolution of the signal over time at different positions. We observe that the two-body lifetime of the signal at a displacement of $d$=1.4mm from the trap center is less than half of the corresponding value measured at $d$=0 (Fig. 4B). This finding indicates a non-uniform evolution of the density distribution in the trap. Specifically, it appears that the density decays faster near the trap's edge than near its center.

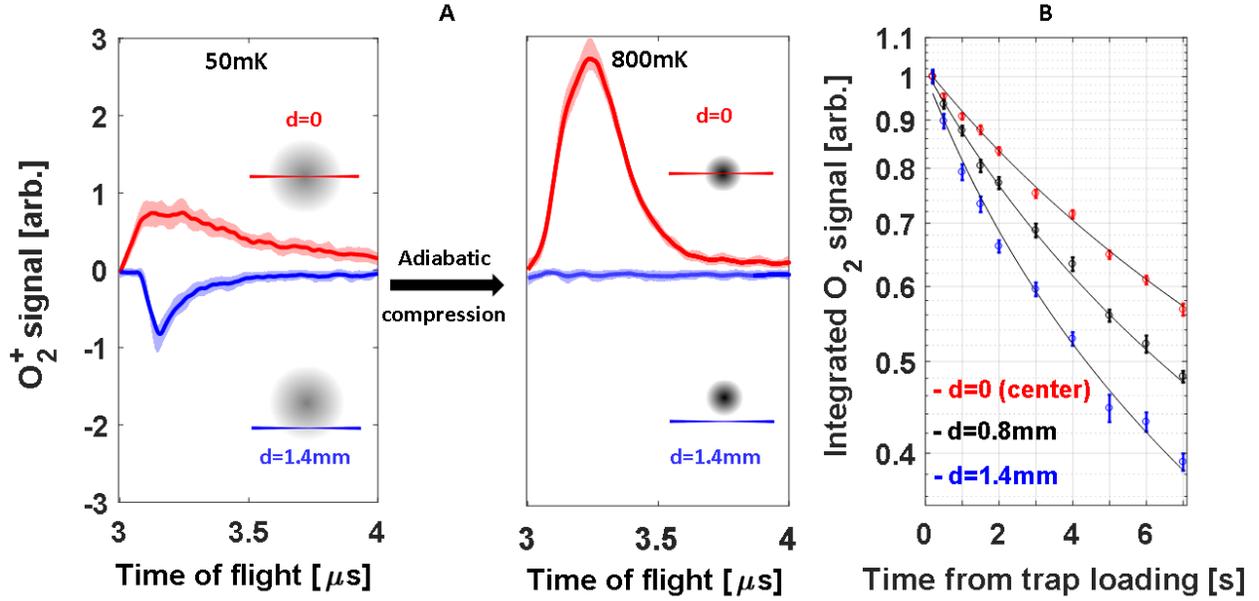

**Figure 4:** Spatial probing of the trap. (A) Compression of the trap from 50mK to 800mK shows a strong gain in the signal measured through the trap center (top red), as observed in the time-of-flight traces of ions arriving on the MCP. In contrast, the signal observed at a transverse displacement of $d$=1.4mm is completely lost after compression (bottom blue). The shaded area indicates the standard deviation of the ensemble average. (B) The two-body lifetime $\tau_{2B}$ decreases when the signal is measured at increasing displacement from the trap center, from 9.1s at $d$=0, to 6.3s at $d$=0.8mm, to 4.3s at $d$=1.4mm, indicating a significant contribution of evaporative losses due to elastic collisions.

To better understand this positional variation in the density evolution, we implement a direct-simulation Monte Carlo (DSMC) approach to model the classical dynamics of the molecules in the trap[42]. At each time step, collision partners are sampled from the velocity distribution for every volume element in the trap. The collision probability is a function of relative velocity and cross-section. In our simulation the total collision cross-section is taken as a constant; once a collision between two particles occurs, the probability of an elastic or inelastic outcome is given according to a pre-defined ratio between the respective cross-sections, $\varepsilon = \sigma_e/\sigma_i$. Inelastic collisions are assumed to lead to loss of both participating particles. Additionally, simulated particles can be lost due to elastic collisions with sufficiently high relative velocities that lead to ejection from the trap. Losses due to background collisions are neglected due to their longer time scale. Additional details regarding this simulation are given in the Supplementary Material.

With this simulation, we calculate the evolution of the column-integrated density at different displacements from the trap center. For early times, it can be shown that the ratio of the two-body lifetimes between any two displacements is only a function of $\varepsilon$ (see Supplementary Material). We can thus use the ratio between the two-body lifetimes measured at $d$=1.4mm and at $d$=0, $\eta_{2B(d=1.4mm)} = \tau_{2B(d=1.4)}/\tau_{2B(d=0)}$, as an indicator for the elastic to inelastic ratio. Figure 5 presents results of simulations performed with different values of $\varepsilon$. When inelastic collisions dominate, the loss rate is faster at the trap center where the initial trapped density is higher. For $\varepsilon>1$, where elastic collisions are preferred, the decay closer to the trap edge become increasingly faster due to an increase in evaporative losses. Above $\varepsilon\approx20$, elastic collisions push a sufficient amount of lower energy particles inwards to cause an initial increase in signal as measured through the trap center, leading to a negative value of $\tau_{2B(d=0)}$.

Applying these results to our decay rate measurements, we find that $\eta_{2B(d=1.4mm)}$ is in the range of 0.3 to 0.6. From Fig. 5 we can therefore conclude that the ratio of elastic to inelastic cross-sections for oxygen-oxygen collisions at 800mK is on the order of $\varepsilon=4$ to 8. These values are in good agreement with a theoretical study[41] of the cross-sections at low temperatures, which predicts that the ratio varies from about 1 at 10mK to almost 10 at 1K.

Finally, with a value for the elastic to inelastic ratio, we can derive the total cross-section. For $\varepsilon=6$, we find in the simulation that $(\tau_{2B(d=0)})^{-1}$ is about one quarter of the mean collision rate, indicating that the initial collision rate in the 800mK trap is about 0.5Hz. As the decelerator can load particles at densities on the order of $10^{10}$ per cubic cm [37], we estimate the average total cross-section for $O_2$-$O_2$ collisions in this trap to be on the order of 500Å$^2$. This estimate is also in reasonable agreement with a theoretical calculation[41].

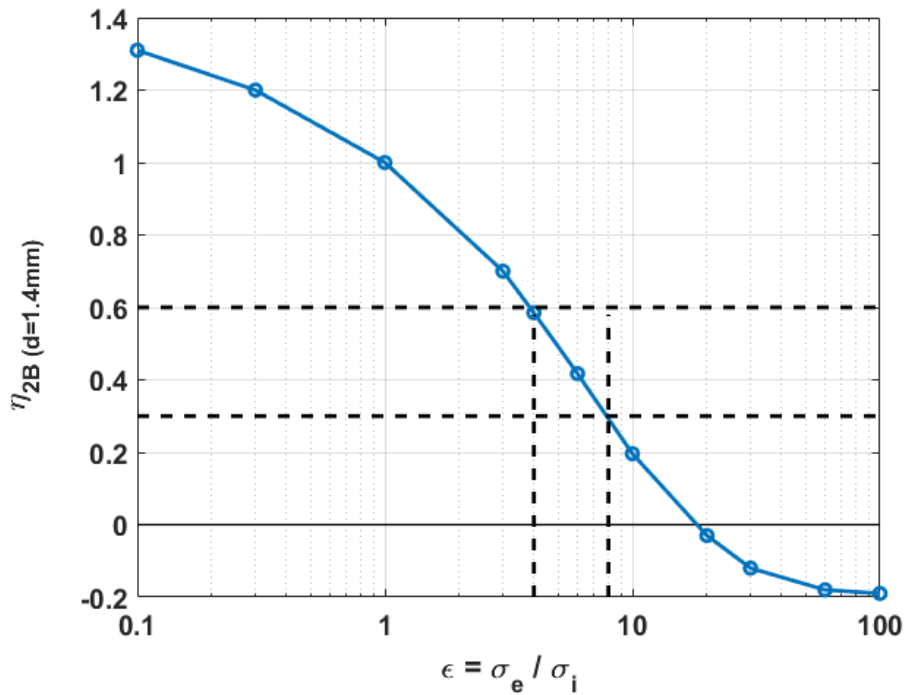

**Figure 5:** Effect of collisional properties on the spatial variation of the decay rate. Using a DSMC simulation, we evaluate how $\varepsilon$, the ratio of elastic to inelastic collision cross-sections, affects the two-body lifetime measured at different displacements $d$ from the trap center. The solid curve represents the expected ratio between the value of the lifetime at $d=1.4$mm and $d=0$ for different values of $\varepsilon$. The experimentally measured ratio lies between 0.3 and 0.6 (horizontal dashed lines), corresponding to an elastic to inelastic ratio of 4 to 8 (vertical dashed lines).

**Discussion**

The novel trapping approach we have presented enables the observation of cold bimolecular collisions due to the combination of high trapping density and long interaction times. Compared to the beam based methods, where the interaction is limited to several tens of microseconds, we increase the probing time by six orders of magnitude, opening up a variety of options for collision studies that are currently unattainable.

One example is the investigation of the effects of symmetry in molecule-molecule collisions. For oxygen, the most abundant $^{16}O^{16}O$ molecule can be replaced with $^{17}O^{17}O$, which has only even-numbered molecular angular momentum states due to the nuclear symmetry properties, or with the heteronuclear $^{16}O^{18}O$, which possesses all possible rotational states.

Another option is the study of cold reactions, made possible by the ability to co-load mixtures of paramagnetic species into the trap. For example, a reaction of Li and $O_2$ may be studied by optically exciting the Li into the reactive $^2P_{3/2}$ state. Similarly, co-decelerating atomic carbon instead of lithium would enable studying the C-$O_2$ barrierless reaction[43]. By tuning the trap depth, we expect to measure the thermally averaged rate coefficients from several tens of millikelvins up to about 1K.

The ability to tune the trap depth is also convenient for attempting forced evaporative cooling. Lowering the trap depth will induce loss of high-energy particles from the trap. The ensemble can then be brought to thermal equilibrium at a lower temperature by raising back the trap gradient, thus increasing the collision rate. Repeating this sequence with decreasingly lower thresholds will lead to a colder ensemble of molecules. To eventually reach the ultracold regime, and possibly quantum degeneracy, would require a sufficiently high ratio of elastic to inelastic cross-sections, as only the elastic fraction contributes to the thermalization rate and the escape of entropy through evaporation, while inelastic collisions reduce the phase-space density.

For oxygen molecules in our magnetic trap, we estimated a ratio of elastic to inelastic cross sections of about 6, a value considerably lower than the typical values required for efficient evaporative cooling. However, calculations[41] predict that other isotopes, such as $^{17}O^{17}O$, may have better collisional properties at low temperatures due to weaker anisotropy of the molecular potential energy surfaces. Furthermore, the generality of the deceleration and trapping scheme will allow us to test the feasibility for evaporative cooling of other paramagnetic radicals, such as NH, $NH_2$, OH and $CH_3$.

Finally, our experiment enables trapping of molecules amenable to laser cooling, without the need for Zeeman slowing, which would otherwise be inefficient for molecules due to the requirement for thousands of closed scattering cycles. The magnetic field in the high-temperature superconducting trap can be relaxed or switched off in very short times, enabling the initiation of a magneto-optical trap.